\documentclass[aip,jcp,amsmath,amssymb,floatfix,reprint,twocolumn]{revtex4-1}
\usepackage{amssymb}
\usepackage{amsmath}
\usepackage{graphicx}
\usepackage{amsbsy}
\usepackage{color}
\usepackage{bm}

\begin{document}
\title{From rods to helices: evidence of a screw-like nematic phase}

\author{Hima Bindu Kolli} 
\affiliation{Dipartimento di Scienze Molecolari e Nanosistemi, 
Universit\`{a} Ca' Foscari di Venezia,
Dorsoduro 2137, 30123 Venezia, Italy}

\author{Elisa Frezza} 
\affiliation{Dipartimento di Scienze Chimiche, 
Universit\`{a} di Padova, via F. Marzolo 1, 35131 Padova, Italy}

\author{Giorgio Cinacchi} 
\email{giorgio.cinacchi@uam.es}
\affiliation{
Departamento de F\'{i}sica Te\'{o}rica de la Materia Condensada and
Instituto de F\'{i}sica de la Materia Condensada, 
Universidad Aut\'{o}noma de Madrid,
Campus de Cantoblanco, 28049 Madrid, Spain
}
\author{Alberta Ferrarini} 
\email{alberta.ferrarini@unipd.it}
\affiliation{Dipartimento di Scienze Chimiche, 
Universit\`{a} di Padova, via F. Marzolo 1, 35131 Padova, Italy}

\author{Achille Giacometti} 
\email{achille.giacometti@unive.it}
\affiliation{Dipartimento di Scienze Molecolari e Nanosistemi, 
Universit\`{a} Ca' Foscari di Venezia,
Dorsoduro 2137, 30123 Venezia, Italy}

\author{Toby S. Hudson}
\email{toby.hudson@sydney.edu.au} 
\affiliation{School of Chemistry, University of Sydney, NSW 2006, Australia}

\date{\today}

\begin{abstract}
Evidence of a special chiral nematic phase is provided 
using numerical simulation and Onsager theory
for systems of hard helical particles.
This phase appears at 
the high density end of the nematic phase, 
when helices are well aligned, and 
is characterized by the C$_2$ symmetry axes of the helices spiraling around 
the nematic director with periodicity equal to the particle pitch. 
This coupling between translational and rotational degrees of freedom allows 
a more efficient packing and hence an increase of translational entropy. 
Suitable order parameters and correlation functions are introduced to identify this screw-like phase,
whose main features are then studied as a function of radius and pitch of the helical particles.
Our study highlights the physical mechanism underlying a similar ordering observed 
in  colloidal helical flagella 
[E. Barry et al. \textit{Phys. Rev. Lett.} \textbf{96}, 018305 (2006)] and
raises the question of whether it could be observed in other helical particle systems, 
such as DNA, at sufficiently high densities. 
\end{abstract}
\maketitle
High density solutions of helical polynucleotides and polypeptides are known 
to form liquid--crystal phases that have 
important physico--biological consequences \cite{list,review1}.  
The simplest of these phases is the nematic (N)  where 
the helices, as elongated particles, have 
their long axes preferentially aligned along a common fixed direction $\widehat{\mathbf{n}}$, 
whereas their centers of mass are homogeneously distributed in space \cite{dG}.
The intrinsically chiral character of the helical constituents may translate into
a chiral organization, the cholesteric phase \cite{dG}, in which 
$\widehat{\mathbf{n}}$ twists around a perpendicular axis 
with a periodicity in the $\mu$m range and 
whose value finely reflects the nm-ranged molecular structure and intermolecular interactions.
This organization can be traced back to 
a preferential mutual twist of 
nearby particle long axes \cite{twist}.

\begin{figure}[htbp]
\begin{center}
\includegraphics[width=2.5in]{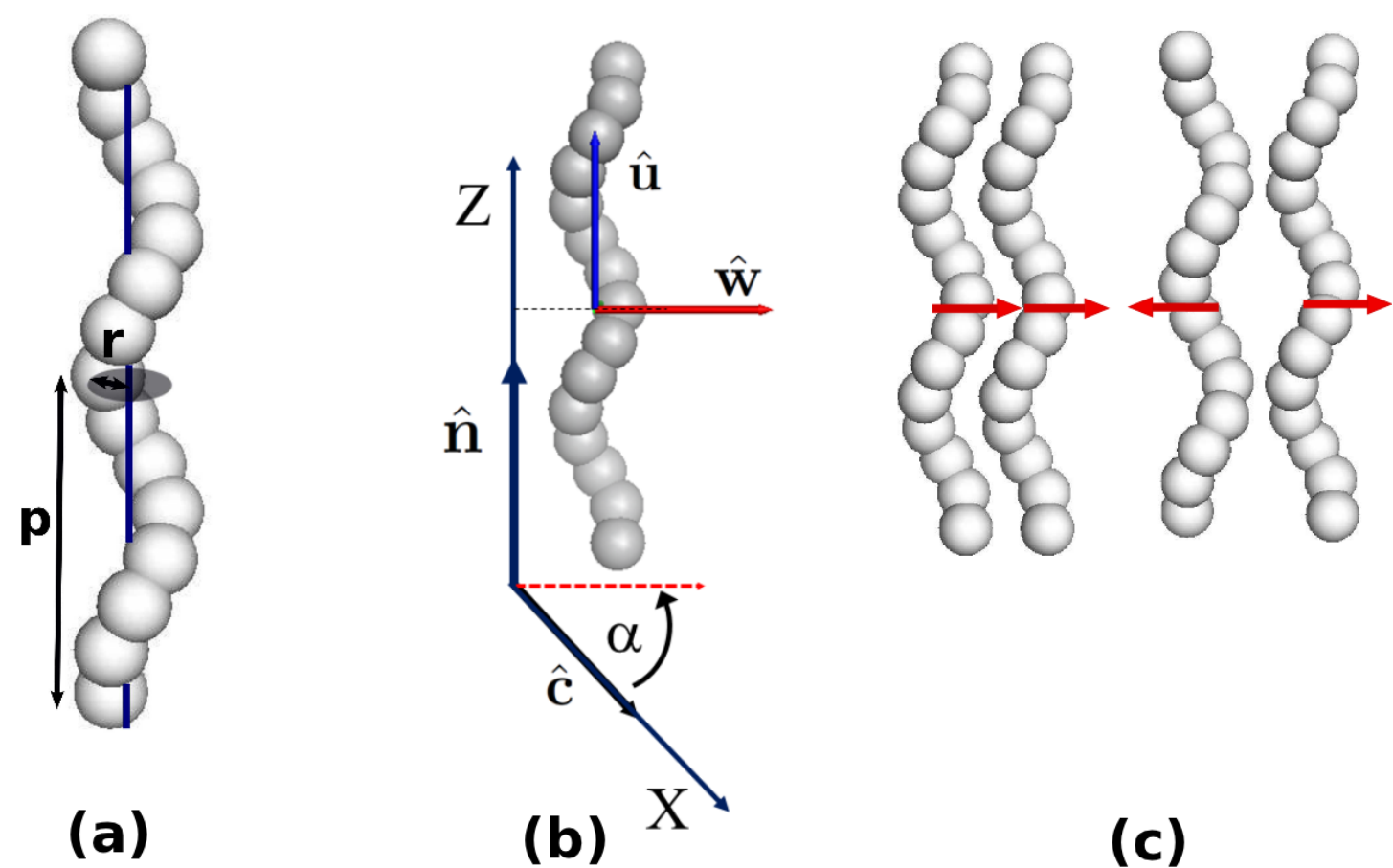}
\end{center}  
\caption{(a) Model helix of radius $r$ and pitch $p$ made of 
15 partially fused hard spheres of diameter $D$ and contour length $L$=10$D$. 
(b) Helix with arrows showing the unit vectors $\widehat{\mathbf{u}}$ and  $\widehat{\mathbf{w}}$, 
parallel to its long axis and to its twofold symmetry axis, respectively; 
$(\widehat{\mathbf{n}}, \widehat{\mathbf{c}})$ are unit vectors parallel 
to the main phase director and the minor phase director at a given position, respectively;
$X$, $Z$ are the axes of the laboratory frame, with  
$\alpha$ the angle between $\widehat{\mathbf{w}}$ and the $X$ axis. 
(c) Pairs of helices in phase and antiphase.}

\label{fig:fig1}
\end{figure}

In this work, 
using Monte Carlo (MC) simulations complemented by Onsager theory, 
we provide convincing evidence of 
the existence of a different chiral nematic phase, 
originating from the specific helical shape of the particles and 
stable against other possible phases within
a specific range of densities, dependent on 
the radius $r$ and the pitch $p$ of the constituent helices [Fig.\ref{fig:fig1}(a)].

Similarly to the cholesteric, this phase is still nematic in that 
helices are homogeneously distributed and mobile with 
their long axis $\widehat{\mathbf{u}}$ preferentially oriented along 
the main director $\widehat{\mathbf{n}}$. 
Differently from the cholesteric, in this new organization 
it is the  short axes $\widehat {\mathbf{w}}$, 
parallel to the twofold (C$_2$) 
symmetry  axes  of the helices, that become long-range correlated and 
preferentially oriented along a second common director, 
$\widehat{\mathbf{c}}$, that in turn spirals around $\widehat{\mathbf{n}}$ with 
a periodicity equal to the helix pitch. 
In the following, we will refer to this special nematic phase as 
\textsl{screw-like} and denote it by N$^{*}_s$. 
Fig.\ref{fig:fig1}(b) shows the unit vectors  $\widehat{\mathbf{n}}$ and 
$\widehat{\mathbf{c}}$, parallel to the phase directors, 
as well as $\widehat{\mathbf{u}}$ and $\widehat{\mathbf{w}}$, 
 which are attached to the helix and are parallel 
to its long axis and to the C$_2$ symmetry axis, respectively.  
In the N$^{*}_s$ phase, 
there is a difference between the longitudinal and the transversal order: 
the former is non-polar 
(there is a $\widehat{\mathbf{n}}\leftrightarrow-\widehat{\mathbf{n}}$ symmetry), 
whereas the latter is polar 
( $\widehat{\mathbf{c}}$ and -$\widehat{\mathbf{c}}$ are not equivalent) and spiraling. 
Thus this new phase is profoundly different from  the biaxial nematic phase, 
where  the minor director is both uniform and nonpolar. 

A transition from the isotropic (I) to the N$^{*}_s$ phase as density increases 
has been observed in colloidal suspensions of 
helical flagella by Barry \textit{et al.} \cite{flagella}, 
using  polarizing and differential interference contrast microscopy,
combined with experiments on single-particle dynamics. 
 A striped birefringent pattern was observed consistent with a picture where
the local tangent to each helix, tilted with respect to $\widehat{\mathbf{n}}$,
is rotating in a conical way. By analogy with a similar behavior occurring for the cholesteric phase
in the presence of an external field parallel to the twist axis  \cite{Meyer68,Kamien92},
this phase was denoted as conical,     
although the physical underlying mechanism and detailed structure is different. 

Consider a pair of helices locally in phase 
contrasted with the case where they are in antiphase, 
as shown in Fig.\ref{fig:fig1}(c). 
While in the latter case both helices can freely rotate about their $\widehat{\mathbf{u}}$ axis, 
effectively behaving as independent cylinders,
this is no longer the case for the in phase case, 
where one of the two helices has to perform a specific additional translation along 
its $\widehat{\mathbf{u}}$ axis in order to rotate about the other fixed helix. 
As elaborated in details below, at sufficiently high densities, this local roto-translation coupling propagates to the whole system and 
originates the N$^{*}_s$ phase.

We modeled the helices as a set of 
15 partially fused hard spheres of diameter $D$, 
our unit of length,  
rigidly arranged in a helical fashion with a given contour length $L=10D$
[Fig.\ref{fig:fig1}(a)]. Hence 
different helix morphologies can be achieved upon changing 
$r$ and $p$ \cite{helix1}, 
as in the experiments on flagella
\cite{flagella}. 
We then performed MC isobaric-isothermal ($NPT$) numerical simulations \cite{Allen},  
on systems of $N$, typically between $900$ and $2000$, such helices, 
at many values of pressure $P$, measured in reduced units $P^{*}=PD^3/k_BT$,
$k_B$ being the Boltzmann constant and $T$ the temperature.
Our simulations were organized in cycles, each consisting, on average, of
$N/2$ attempts to translate a randomly selected particle, $N/2$ trial rotational moves and 
an attempt to modify shape and volume of the triclinic computational box.
Periodic boundary conditions were applied, as these are appropriate in the present case. 
Initial configurations were taken either
as a low density or a highly ordered compact configuration.
Typically,  
$3-4 \times 10^6$ equilibration MC cycles were followed by 
additional $2 \times 10^6$ production MC cycles 
to collect statistics on various quantities.

\begin{figure}[htbp]
\begin{center}
\includegraphics[width=2.5in, angle=-90]{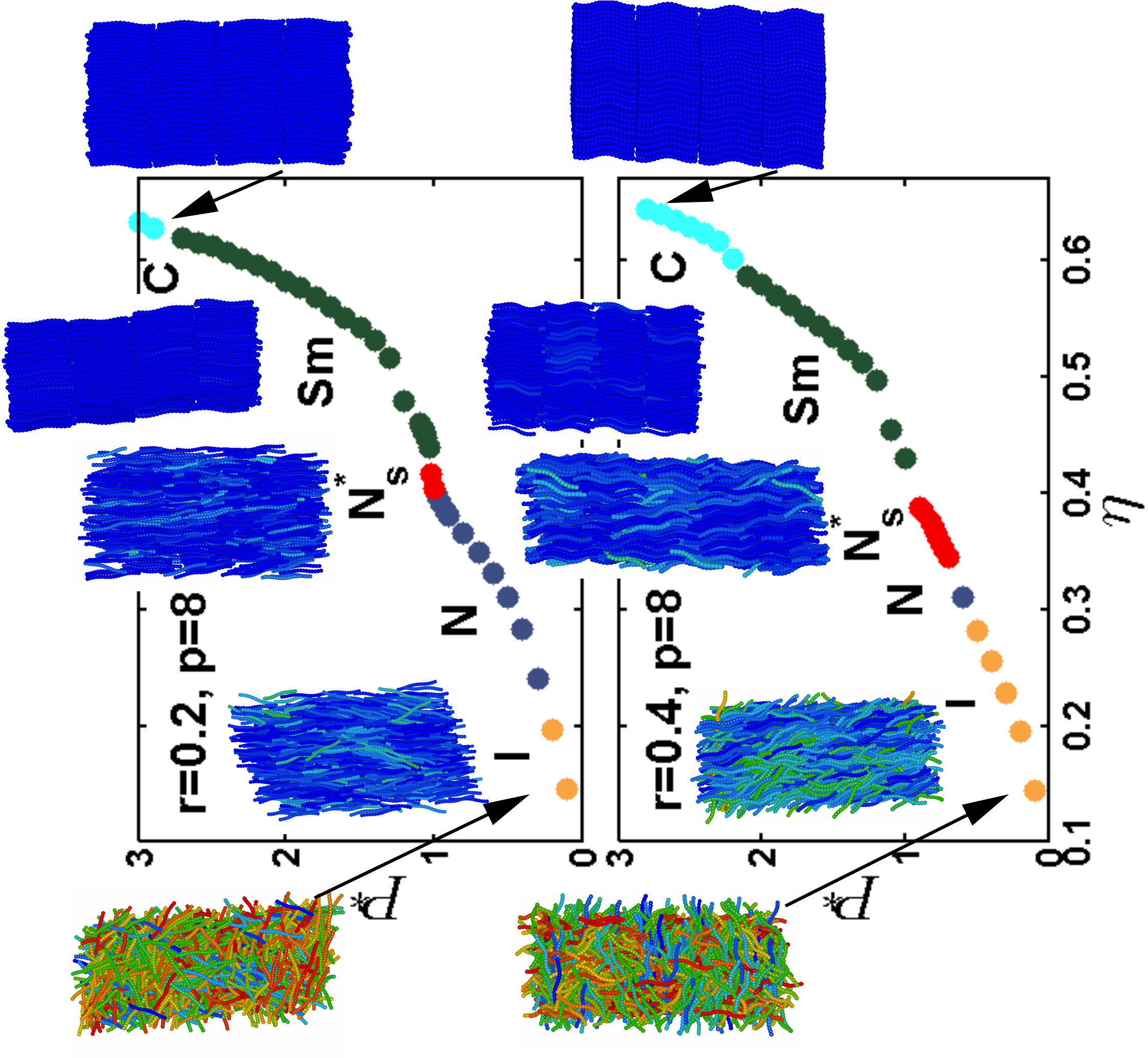} 
\end{center} 
\caption{
 $P^{*}$ as a function of  $\eta$ 
for systems with $r=0.2,0.4$ and $p=8$.
Representative snapshots of each phase 
(I=Isotropic, N=Nematic, N$^{*}_s$=screw-like, Sm=Smectic, C=compact) 
are also given in each panel. 
Here different colors represent 
different orientations of the helix axis $\widehat{\mathbf{u}}$ with respect to the direction $\widehat{\mathbf{n}}$.
}
\label{fig:fig2}
\end{figure}

\begin{figure}[htbp]
\begin{center}
\includegraphics[width=3.0in]{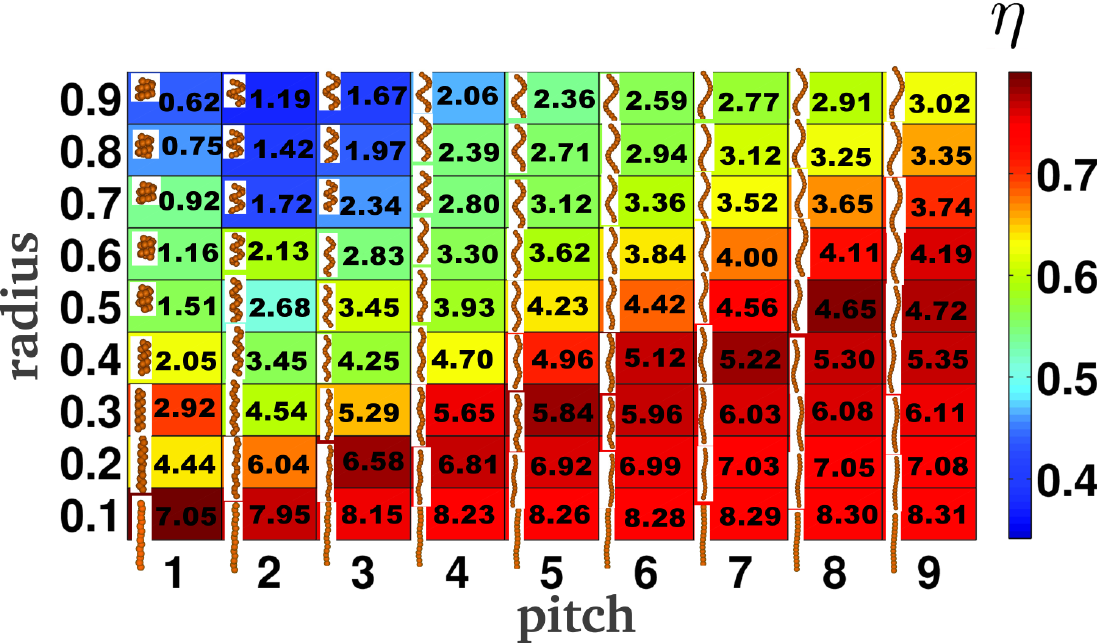}  
\end{center}  
\caption{Color map of the maximal volume fraction, $\eta_{\text{max}}$, 
achievable by a given helix 
as a function of its radius $r$ and pitch $p$.
The color code is from dark red (high packing) to dark blue (low packing), 
The values on the side of each helix side is its effective aspect ratio, $L/(2r+D)$.                
The cartoon on the bottom left of each $r$,$p$ pair shows the corresponding helix morphology. }
\label{fig:fig3}
\end{figure}

Fig.\ref{fig:fig2} shows, in the $P^*$--volume fraction ($\eta=\rho v_0$,
with $\rho$ the number density and $v_0$ the helix volume) plane, 
the MC results for the representative cases with $r=0.2$ and $r=0.4$ and the same value of $p=8$. 
Points labeled I, N, Sm correspond to the 
isotropic,  ordinary nematic and smectic phases, respectively, as
identified by the usual nematic 
$\langle P_2 \rangle=(3\langle ({\widehat{\mathbf{u}}}\cdot {\widehat{\mathbf{n}}})^2 \rangle -1)/2$ and 
smectic 
$\tau_1= \left \vert \left \langle \exp\left(i{2\pi Z}/{d} \right) \right \rangle \right \vert$ ($d$ being the layer spacing, and $Z$ being
the position along the $\widehat{\mathbf{n}}$ axis) 
order parameters \cite{dG}, as well as appropriate correlation functions. 
Points identified by C correspond to compact phases. 
Note that, unlike the case of spherocylinders \cite{Bolhuis97}, 
the high-density phase diagram of helices is not known,  
and constitutes an interesting property on its own right. We determined the maximum packing configuration
by adapting a methodology proposed in Ref.\cite{Hudson}, that hinges on an annealing reorganization scheme 
for a unit cell toward the most compact configuration, 
in the approximation where we consider a single layer of helices.
The results of this analysis are summarized in the color map of Fig.\ref{fig:fig3} 
displaying the largest obtained volume fraction as a function of $r$ and  $p$ of the helices. 
Interestingly, while there exists a large variation of the maximal packing, 
depending on the helix morphology, similar values can be achieved for different $r$, $p$ pairs. 
The high density state points in the phase diagram of Fig.\ref{fig:fig2} were obtained using these maximal
packing configurations as initial conditions, upon applying the appropriate pressure until equilibration.
Points denoted by C in Fig.\ref{fig:fig2} are then associated with highly ordered configuration compatible with
solid-like ordering.
Finally, points identified by N$^{*}_{s}$ in Fig.\ref{fig:fig2} correspond to the special N phase with screw-like order. 
These are the central results of the present study and 
require a special set of correlation functions and order parameters to be fully characterized.
One key orientational correlation function is
$g_{1,\|}^{\widehat{\mathbf{w}}}(R_{\|}) = \langle \widehat{\mathbf{w}}_i \cdot \widehat{\mathbf{w}}_j  \rangle_{(R_{\|})}$,
where  $\widehat{\mathbf{w}}_i$ is a unit vector along the C$_2$ symmetry axis of helix $i$ (Fig.\ref{fig:fig1} (b)), and the
subscript
$R_{\|}=\mathbf{R}_{ij} \cdot \widehat{\mathbf{n}}$ means that the average is restricted to  
pairs of helices $i$ and $j$ with a specific $R_{\|}$, the projection of the interparticle separation
$\mathbf{R}_{ij}$ along the $\widehat{\mathbf{n}}$ axis. 
Thus $g_{1,\|}^{\widehat{\mathbf{w}}}(R_{\|})$ probes the polar correlation between 
the C$_2$ symmetry axes of two helices as a function of their distance projected along the main director.
\begin{figure}[htbp]
\begin{center}
\includegraphics[width=3.0in, angle=-90]{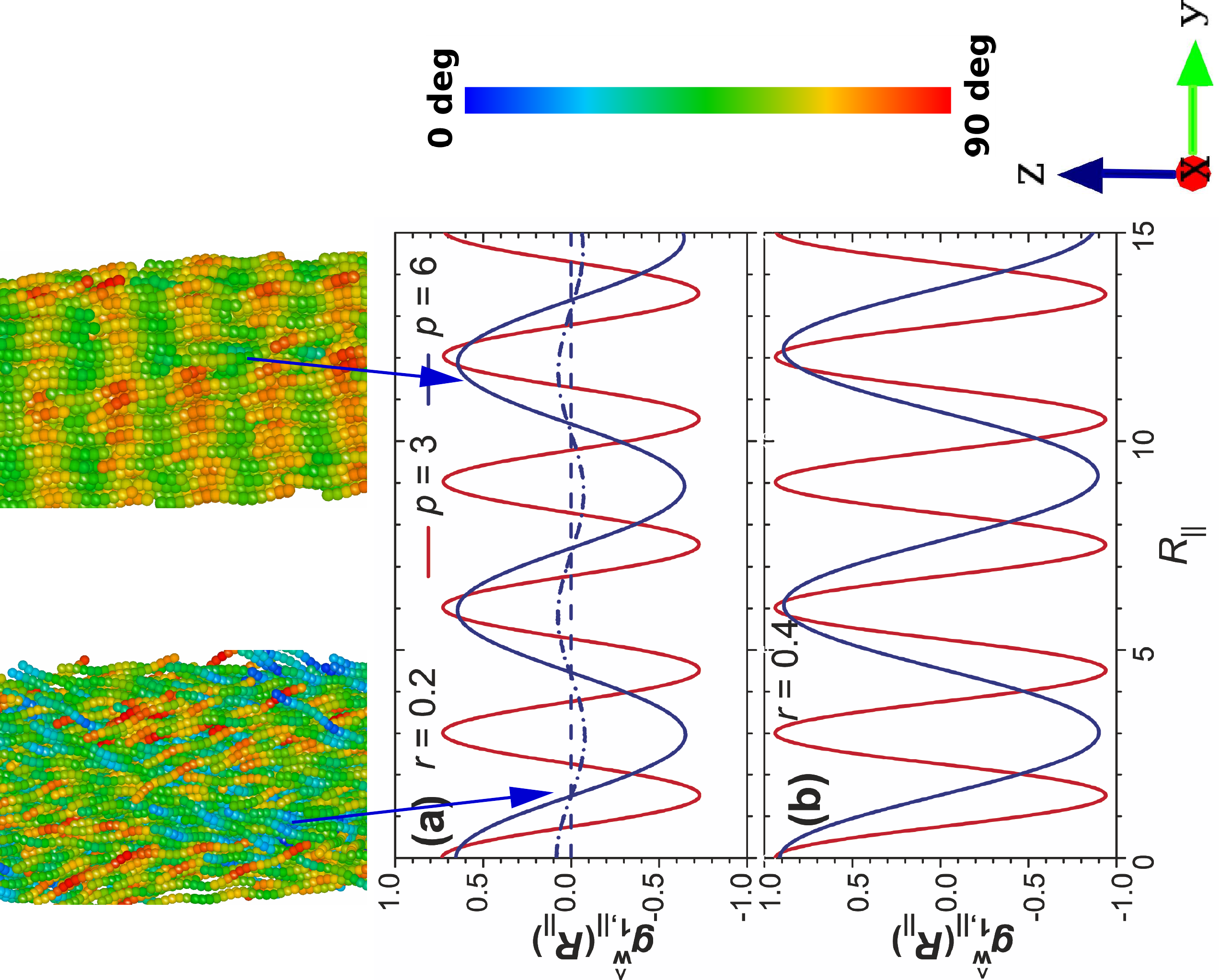} 
\end{center}  
\caption{The correlation function $g_{1,\|}^{\widehat{\mathbf{w}}}(R_{\|})$. 
(a) $r=0.2$  and  $p=3$ (red) at $\eta \simeq 0.38$, and $p=6$ (blue) at 
$\eta\simeq 0.40$ (solid), $\eta\simeq 0.36$ (dashed dotted) and $\eta\simeq 0.27$ (dashed). 
On top of this panel, two representative snapshots of the cases $\eta\simeq 0.27$ (left, standard nematic) and  
$\eta\simeq 0.40$ (right,  screw-like nematic) are depicted,
giving a visual difference between the two phases. 
Note that in this case, at variance with Fig.\ref{fig:fig2}, 
what is color coded is the local tangent of each helix and the reference
axis corresponding to 0 degrees is the straight line x=y=z.
(b) $r=0.4$ and $p=3$ at $\eta \simeq 0.41$ and $p=6$ at $\eta \simeq 0.40$.} 
\label{fig:fig4}
\end{figure}
\begin{figure}[htbp]
\begin{center}
   \includegraphics[width=3.0in]{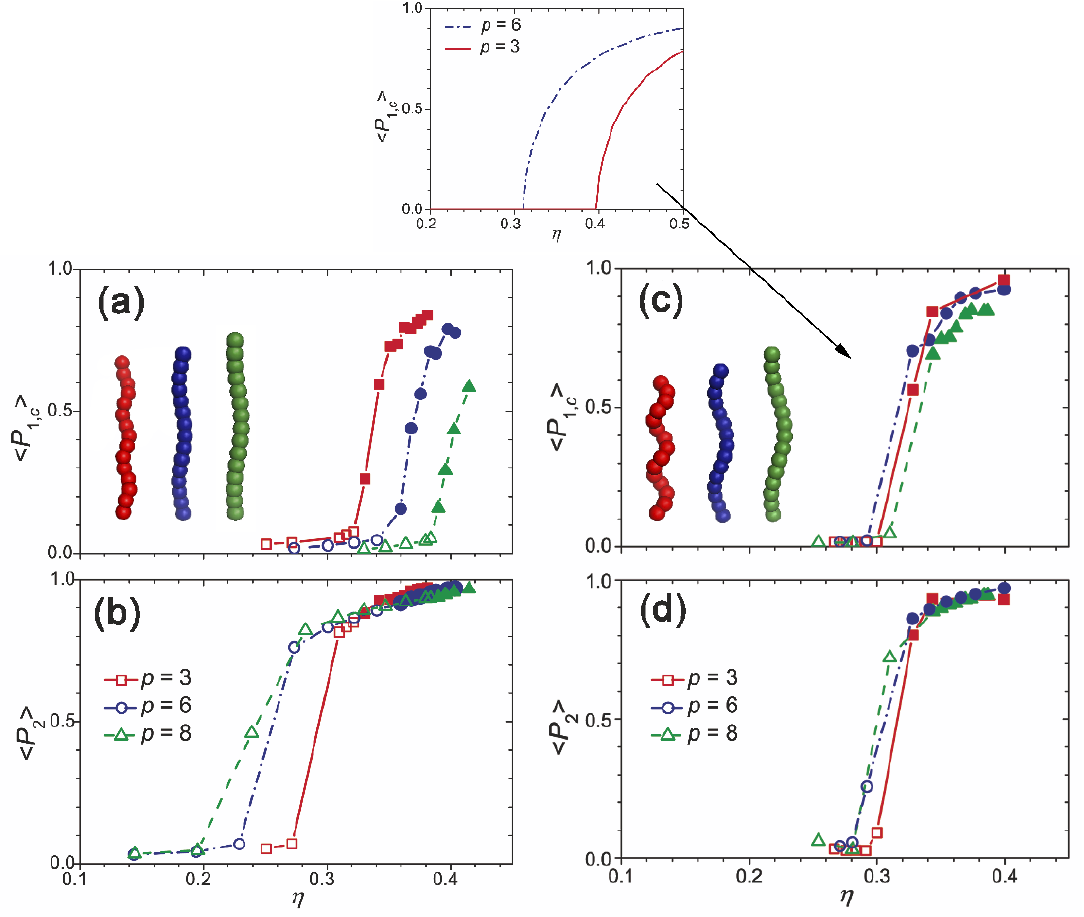}  
\end{center}  
\caption{The order parameters $\langle P_{2} \rangle$ and 
$\langle P_{1,c} \rangle$ as a function of $\eta$, 
in the case $r=0.2$ (a,b) and $r=0.4$ (c,d) and 
different values of $p=3;6;8$. 
Solid symbols are only used for $N^*_s$ phase. The inset depicts the result from Onsager theory.} 
\label{fig:fig5}
\end{figure}
Fig.\ref{fig:fig4} shows this correlation function calculated for helices with 
$r=0.2$ (a) and $0.4$ (b) and  $p=3$ and $6$,  at different values of $\eta$.
For both radii, a sinusoidal structure with a periodicity equal to $p$ is clearly visible. 
It persists with a constant amplitude at long interparticle distances. 
This behavior reflects the helical correlation of the  $\widehat{\mathbf{w}}$'s
along $\widehat{\mathbf{n}}$ and 
is the unambiguous signature of the screw-like ordering.
The onset of this ordering with increasing density is clearly visible in Fig. \ref{fig:fig4}(a). 
At $\eta \simeq0.27$,  $g_{1,\|}^{\widehat{\mathbf{w}}}(R_{\|})$ is nearly zero everywhere, 
showing the lack of transversal correlation between helices. A glance to the corresponding snapshot
supports this interpretation. Note that here, at variance with the case of Fig.\ref{fig:fig2}, different colors
represent different directions of the local tangent to each helix, in analogy with experiment of Ref.\cite{flagella}.  
At $\eta \simeq 0.36$ a small amplitude oscillation can be distinguished, 
which continuously grows up with increasing density, 
toward the condition of perfect ordering, 
where $g_{1,\|}^{\widehat{\mathbf{w}}}(R_{\|})$ would oscillate between $\pm$1. 
At $\eta \simeq 0.40$ the correlation is fully developed as indicated by its significant oscillation and
reflected in the stripes appearing in the corresponding snapshot. 
In Fig. \ref{fig:fig4}(b) only sinusoidal curves are present as for $r=0.4$
the N phase is either absent altogether or surviving in a very narrow range of $\eta$. 
Additional insights on the onset of the N$^{*}_s$ phase can be found by considering
the lowest--rank order parameter 
$\langle P_{1,c} \rangle = \langle \widehat{\mathbf{w}}\cdot \widehat{\mathbf{c}} \rangle$,
related to that proposed in Ref. \cite{Manna} in the framework of 
a theoretical description of a screw-like organization. 
$\left \langle P_{1,c} \right \rangle $  distinguishes the N$^{*}_s$ from the standard N phase, 
as both are characterized by a non-zero value of  $\langle P_2 \rangle$. 
Fig.\ref{fig:fig5}(a,b) shows both $\langle P_2 \rangle$ and  $\langle P_{1,c} \rangle$  as a function of $\eta$, 
for helices with  $r=0.2$ and increasing value of the $p=3$, $6$, and $8$.
As pitch increases, the location of the 
I-N phase transition moves to lower $\eta$, as indicated by the  $\langle P_2 \rangle$ behavior, 
in agreement with results reported earlier \cite{helix1}.
This can be understood in terms of an increase of the  effective
 aspect ratio that tends to stabilize the N phase.
The location of the N to N$^{*}_s$ phase transition instead moves to larger $\eta$ for increasing pitch,
with a significant pitch dependence, and with the  
N$^{*}_s$ phase always occuring at high values of $\langle P_2 \rangle$. 
This can be ascribed to the fact that
the nematic order has first to set in and reach a very high degree
before the C$_2$ symmetry axes start twisting around  $\widehat{\mathbf{n}}$ and
become long-range correlated, enhancing translational entropy.  
The behavior of the $\langle P_{1,c} \rangle$ in the neighborhood of 
the N to N$^{*}_s$ phase transition is suggestive of a second-order phase transition, 
at variance with the first-order character of the usual I-N phase transition. 
The situation is markedly different for the cases with $r=0.4$, depicted in Fig. \ref{fig:fig5}(c,d). 
While there is a similar trend, albeit much less pronounced, 
of the I-N phase transition approximatively shifting toward larger $\eta$ for decreasing $p$,
the seemingly simultaneous rise of both $\langle P_2 \rangle$ and $\langle P_{1,c} \rangle$
is suggestive of a very narrow ordinary N phase or even of
a direct transition from the I to the N$^*_s$ phase.

The onset of the N$^{*}_s$ phase can be further rationalized by an Onsager theory \cite{Onsager}.
Since numerical simulations indicate that the N$^{*}_s$ phase forms at very high values of $\langle P_2\rangle$, 
we here assume perfectly parallel helices ($\langle P_2\rangle=1$).
The single-particle density can be then expressed as a function of 
$\alpha^\prime=\widehat{\mathbf{w}}\cdot \widehat{\mathbf{c}}$,
so that $f(\alpha^\prime)$  is the local orientational distribution function. 
In the N phase the latter is a constant, $f=1/2 \pi$, 
with the normalization condition $\int_0^{2 \pi} d \alpha^\prime f(\alpha^\prime)=1$.
The orientational and excess contribution to the Helmholtz free energy 
is then expressed as a functional of the single-particle density 
${\cal{F}}[f(\alpha,Z)]$, with $Z$ the position of a  particle along  $\hat{\mathbf{n}}$  and 
$\alpha=\alpha^\prime+2 \pi Z/p$ (Fig.\ref{fig:fig1}(b)):
\begin{eqnarray}
\frac{{\cal{F}}}{N k_B T}&=&\int_0^{2 \pi} d \alpha^\prime f(\alpha^\prime)
\ln \left[2 \pi f(\alpha^\prime) \right] \nonumber \\
&+&\frac{\rho}{2} \frac{4-3\eta}{4(1-\eta)^2} \int_0^{2 \pi} d \alpha_1 f(\alpha_1|0) \nonumber \\
&&  \int dZ_{12} \int_0^{2 \pi} d \alpha_2   f(\alpha_2|Z_{12}) a_{excl}(Z_{12},\alpha_1,\alpha_2) \nonumber \\
\label{eq:eq5}
\end{eqnarray}
The first term in right-hand-side of Eq.(\ref{eq:eq5}) represents 
the entropic cost for  the loss of freedom in the azimuthal angle rotation, while 
the second represents the excess free energy 
within the second virial approximation characteristic of Onsager theory.
Here particle positions and orientations are expressed with respect to 
the same (laboratory) reference frame, 
having its origin at the position of the center of the particle 1 and 
the $X$ axis parallel to the $\hat{\mathbf{c}}$ director at this position (Fig.\ref{fig:fig1}(b)).
Vector  ${\bf R}_{12}=(X_{12},Y_{12},Z_{12})$ defines 
the position of particle 2 in this frame,
$\alpha_i$ being the angle between the unit vector $\widehat{\mathbf{w}}$ of  particle $i$ and 
the $X$ axis.
The factor $(4-3\eta)/(4(1-\eta)^2)$ is a correction introduced to account for 
higher virial contributions\cite{parsons-lee} and
$a_{excl}(Z_{12}, \alpha_1,\alpha_2)=-\int dX_{12}\int dY_{12} e_{12}({\bf R}_{12},\alpha_1,\alpha_2)$
(with $e_{12}$ the Mayer function\cite{McQuarrie}) is 
the section of the volume excluded to particle 2 by particle 1 cut by 
a plane normal to $\widehat{\mathbf{n}}$ at $Z=Z_{12}$\cite{spain}. 
The inset of Fig. \ref{fig:fig5} shows the result from Onsager theory for 
the dependence on $\eta$ of the order parameter 
$\langle P_{1,c}\rangle=\int_0^{2\pi} d \alpha^\prime f(\alpha^\prime) \cos(\alpha^\prime)$ 
in the cases $r=$0.4 and  $p$=3 and 6. 
While a quantitative comparison is clearly not possible,
Onsager theory qualitatively agrees with simulation results and clarifies a number of additional issues.
As shown in the inset of Fig.\ref{fig:fig5}, 
the second-order N--N$^*_s$ transition is shifted to a higher $\eta$ for  the helices with a smaller $p$. 
As helices with $p=3$ tend to interpenetrate less than those with $p=6$, the entropic gain
driving the formation of N$^*_s$ phase is correspondingly smaller, and shifts the transition
to higher densities. 
This general pattern is also consistent with the results of the analysis reported in Fig.\ref{fig:fig3} for $r=0.4$.
The small pitch dependence of the I--N--N$^*_s$ transition for $r=0.4$ 
observed in Fig. \ref{fig:fig5}(c,d) can then be
interpreted as a result of two competing effects. 
On the one hand, the shorter effective aspect ratio of the helix associated with smaller $p$ 
tends to push the formation of the N phase at higher densities. 
On the other hand, this is balanced by a larger tendency to develop screw-like ordering. As a result, an almost direct transition
to a N$^*_s$ phase is observed nearly independent on $p$. 

In short, we have found that 
systems of hard helical particles undergo 
a second-order entropy-driven transition from 
an ordinary nematic N to a screw-like nematic  N$^{*}_{s}$ phase at high densities.
We have rationalized the formation of this phase in terms of 
a coupling between translational and rotational degrees of freedom occuring whenever 
there is a sufficiently large interlock of the grooves belonging to neighboring helices.
The N$^{*}_s$ organization is then adopted in order to maximize 
the translational entropy counter-balancing the loss in orientational entropy associated with 
the periodic alignment of the C$_2$ symmetry axes.
By obtaining the full phase diagram of hard helices up to the most compact phases, 
the exact boundaries of the N$^{*}_s$ phase has been determined and
its relative stability analyzed in terms of the helix morphology. 
This screw-like order is specific to helical particles and differs from 
the usual cholesteric order, that requires chirality only.
As these two types of ordering could in principle coexist, 
it would be very interesting to address their compatibility. 
Future work will also include a full characterization of the smectic and compact phases 
specific to helical particles, for which we have already preliminary indications.

Our results provide a theoretical explanation of the I--N$^*_s$ transition observed in helical flagella, 
where polydispersity  most-likely prevents the formation of a smectic phase \cite{flagella}.
This raises the expectation that the same transition could also be observed in other similar systems,
including chiral colloidal particles (e.g. bacteria and viruses), helical (bio)polymers, and 
concentrated DNA solutions, although the much smaller length scales and the specifity of
the interactions involved 
in this latter case may constitute a formidable experimental challenge \cite{Manna}.
Our work, highlights the generality of the entropic mechanism driving the formation of the
N$^{*}_s$ phase in dense systems of helicoidal molecules, which we hope will stir further experimental activities along these
lines.
\begin{acknowledgments}
We are grateful to
C. de Michele, M. Dijkstra, Z. Dogic, D. Frenkel, R. Podgornik, R. van Roij, F. Sciortino, and E. Velasco 
for useful discussions, and to MIUR PRIN-COFIN2010-2011 (contract 2010LKE4CC),  Government of Spain via a Ram\'{o}n y Cajal research fellowship, and
Cooperlink Italy-Australia bilateral agreement for financial support.
\end{acknowledgments}

\end{document}